\documentclass[12pt]{iopart}
\usepackage{booktabs}

\usepackage{graphicx}
\usepackage{longtable}
\usepackage{xcolor}

\newcommand{\cav}{Cavansite}
\newcommand{\dcav}{dehydrated Cavansite}
\newcommand{\pen}{Pentagonite}
\newcommand{\dpen}{dehydrated Pentagonite}

\begin{document}
\title{On the relative abundances of \cav\ and \pen}

\author{Bhalchandra S.\ Pujari}
\address{Department of Scientific Computing Modeling \& Simulation, Savitribai Phule Pune University, Pune 411007, India.}
\ead{bspujari@scms.unipune.ac.in}

\author{Sagar Gehlot}
\address{Department of Scientific Computing Modeling \& Simulation, Savitribai Phule Pune University, Pune 411007, India.}

\author{Mihir Arjunwadkar}
\address{Department of Scientific Computing Modeling \& Simulation, Savitribai Phule Pune University, Pune 411007, India.}

\author{Dilip G.\ Kanhere}
\address{Department of Scientific Computing Modeling \& Simulation, Savitribai Phule Pune University, Pune 411007, India.}

\author{Raymond A.\ Duraiswami}
\address{Department of Geology, Savitribai Phule Pune University, Pune 411007, India.}

\vspace{10pt}
\begin{indented}
\item[]December 2023
\end{indented}

%\addbibresource{biblio.bib}

\begin{abstract}
 \cav\ is a visually stunning blue vanadosilicate mineral with limited occurrences worldwide, whereas \pen\ is a closely related dimorph with similar physical and chemical properties, yet is extremely rare compared to \cav.
 The reasons behind \pen's exceptional rarity remain largely unknown.

 In this study, (a) density functional theory (DFT) is utilized to investigate the electronic structures of \cav\ and \pen\ at ground state and finite pressures; (b) a two-state Boltzmann probability model is then employed to construct a comprehensive phase diagram that reveals the abundance of each species across a wide range of pressure and temperature conditions; and (c) dehydration characteristics of these two minerals are explored.

 The present analysis reveals the key factors that contribute to the relative scarcity of \pen, including differences in structural arrangement and electronic configurations between the two minerals.
 Specifically, it shows that (a) because of the peculiar arrangements of SiO$_4$ polyhedra, \cav\ forms a compact structure (about 2.7\% less in volume) resulting in lower energy; (b) at a temperature of about 650K only about 1\% \pen\ can form; (c) vanadium induces a highly localized state in both of these otherwise large-band-gap insulators resulting in an extremely weak magnetic phase that is unlikely to be observed at any reasonable finite temperature; and (d) water molecules are loosely bound inside the microporous crystals of \cav\ and \pen, suggesting potential applications of these minerals in various technological fields.
\end{abstract}

\vspace{2pc}

\section{Introduction}

Since its discovery\cite{staples1973cavansite} in the 1960s, \cav\ has been a highly coveted mineral due to its spectacular greenish-blue appearance.
It is found in clusters of acicular crystals, with individual crystals being sub-millimeter in size. 
\pen, though chromatically similar, is known for the twining that leads to a near-five-fold-symmetrical structure giving it its name \cite{evans1973crystal,ottens2022secondary}.
Being dimorphs, both have the same chemical formula -- Ca(VO)Si$_4$O$_{10}$$\cdot4$H$_2$O -- and both are orthorhombic crystals with \pen\ having slightly larger (by about 2\% in volume) unit cell.
The blueish color of both the minerals is attributed to vanadium atoms \cite{huang2023revealing}. 
The other common interesting structural character of these dimorphs is the presence of large cavities in their atomic structure due to tetrahedral-pyramidal 
arrangements of vanadosilicate complexes.
Sheets of SiO$_4$ are intertwined in a maze that accommodates calcium and vanadium atoms.
In fact, these are the only known naturally occurring microporous vanadosilicates \cite{DANISI2015257}.
The resulting tunnels are naturally occupied by water and may be dehydrated at higher temperatures \cite{phadke1994thermal,danisi2012situ}.
Such materials are of technological importance due to their ion-exchange, catalytic or sorption applications.
Because vanadium in these complex structures can be in multiple oxidation states they are expected to show catalytic activity; specifically, for selective partial oxidation of ethanol at elevated temperatures \cite{cymes2020catalytic}.
Interestingly, \cav\ was also investigated for bioavailability of vanadium for nitrogen fixation reaction carried out by certain bacteria where it was shown that bacteria may extract vanadium from \cav\ and may incorporate it into cellular biomass \cite{srivastava2023bioavailability,sheng2023mineral}.

Although first discovered in Oregon, USA, arguably best samples of these dimorphs are known to originate from Pune (formerly Poona), India \cite{phadke1994thermal,danisi2012situ}.
Worldwide, these minerals are known to be found at ten locations corresponding to only four basaltic planes; see Table \ref{tab:locations}.
This table, wherever possible, also describes the associated minerals in the corresponding lava flows and shows that the two minerals are found among the upper flows of pahoehoe-type lava.
(Unlike the aa-type flow, a pahoehoe-type lava flow is characterized by a relatively smooth surface.)
The two minerals are formed inside cavities or brecciated surfaces.
We also note that there are at least three sites where \cav\ and \pen\ co-occur, and whereas \pen\ is never found without \cav, the converse is not true.
This gives a clear indication that \cav\ is significantly more abundant than \pen. 
Unfortunately, quantitative field measurements of their relative abundances are absent in field studies due to the rarity of these minerals.
It has been noted, however, that it is nearly impossible to find reasonably large clusters of \pen\ \cite{white2002let}. 
Therefore, considerable uncertainty remains about the nature of the formation processes of \cav\ and \pen.
That they can be found at the same location suggests that the chemical and physical environments responsible for their formation may be similar.
It has been noted that their mineral associations and modes of occurrence are often similar \cite{staples1973cavansite}.
Yet, \pen\ remains significantly rarer.

In this study, density functional theory (DFT) calculations have been used primarily to explore why \pen\ may be so rare.
Formation energies of these minerals deduced thus, and the resulting phase diagram of their relative abundances, suggest why \cav\ remains dominant in the field.
It may be noted that there have been no previous electronic structure investigation on these minerals.

\section{Computational Details}

DFT calculations were performed using the plane-wave basis set and within the Perdew-Burke-Ernzerhof (PBE) generalized gradient approximation \cite{perdew1996generalized}, using Quantum ESPRESSO \cite{giannozzi2009quantum}.
To ensure accuracy, the latest curated pseudopotentials \cite{garrity2014pseudopotentials, dal2014pseudopotentials, kucukbenli2014projector} compiled by Materials Cloud \cite{doi:10.1126/science.aad3000, prandini2018precision} have been used.
Full unconstrained and spin-polarized unit cell optimizations were carried out starting from experimentally reported unit cells consisting of four chemical formula units, resulting in total of 116 atoms.
To ascertain accuracy further, a rather high kinetic-energy cutoff of 950 eV was used for the plane wave basis set, and an automatic 3$\times$3$\times$3 Monkhorst-Pack grid was used to sample the $k$-space during unit cell optimizations.
For subsequent non-self-consistent calculations, a finer $6\times6\times6$ grid was used, and for band structure calculations, 30 $k$-points were used along each symmetry line in Brillouin zone.
Finally, DFT-D3 treatment of Grimme et al.\ \cite{grimmedft} was used to account for van der Waals correction which, as it turns out, plays a vital role in determining lattice structures; this is detailed later in this paper.

\section{Results and discussion}

\subsection{Electronic structure} 

\begin{figure}
\centering
 \includegraphics[width=.34\textwidth]{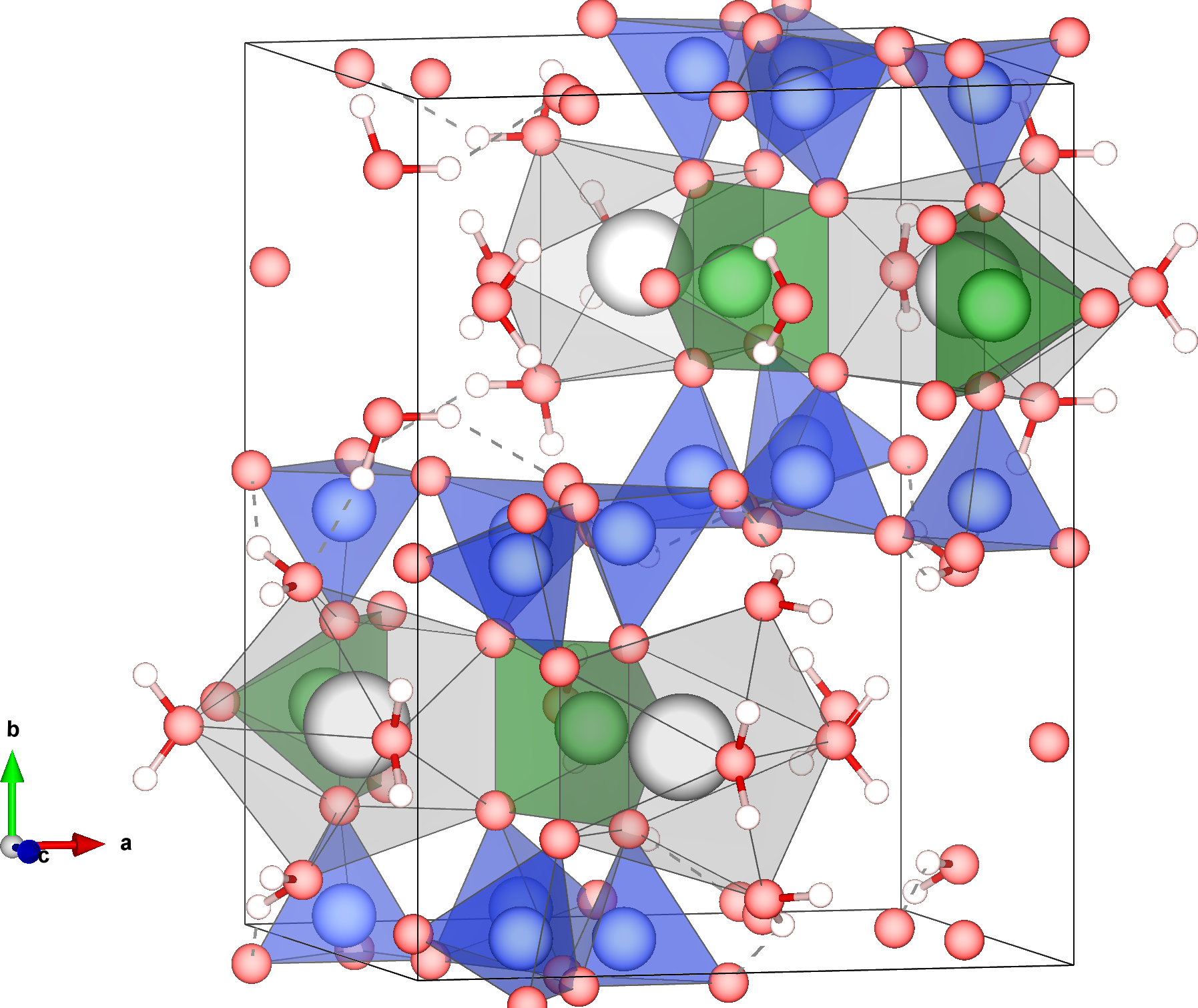}
 \includegraphics[width=.34\textwidth]{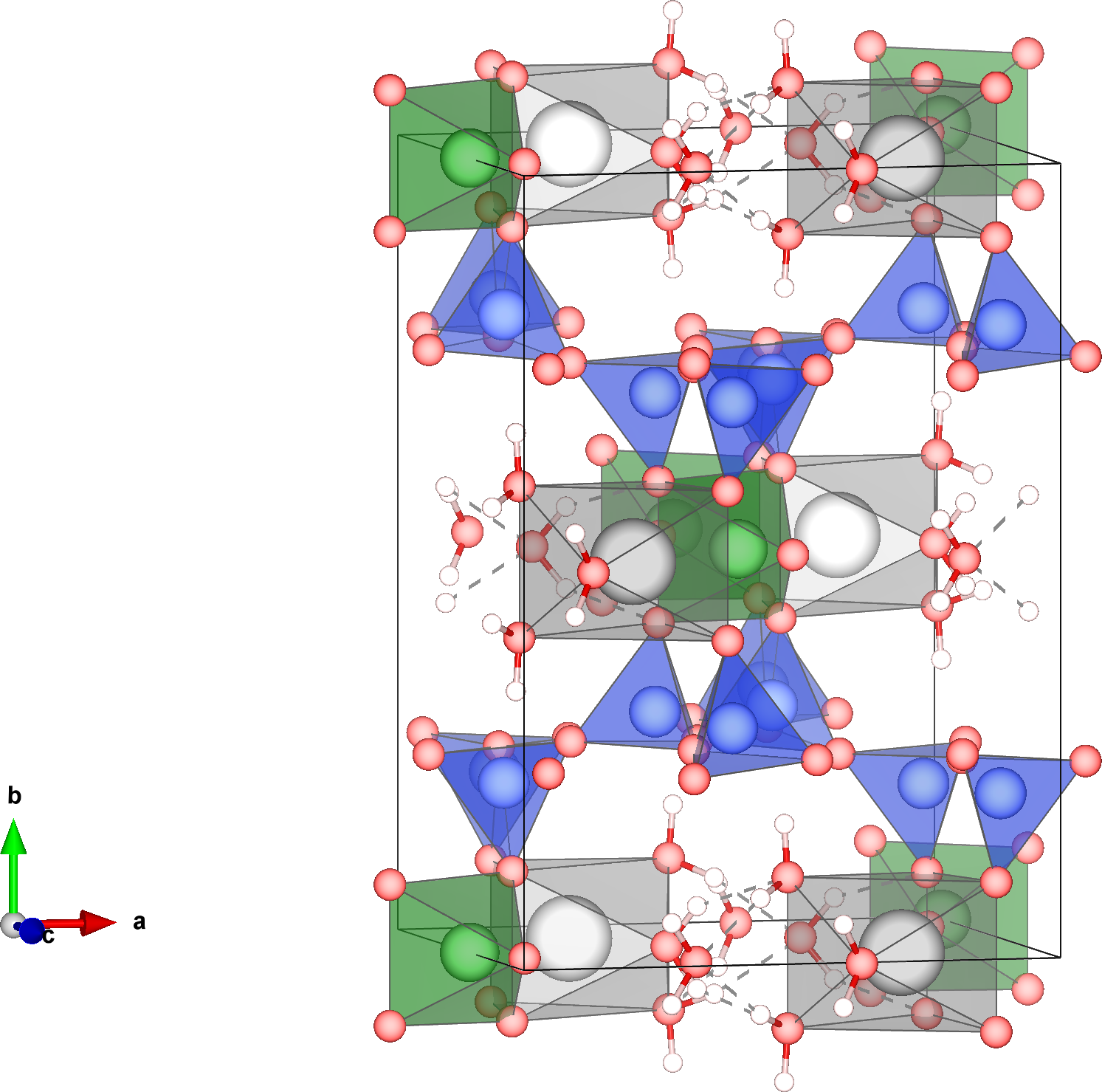}\\
 \centerline{(a) \hskip .34\textwidth (b)} ~\\
 \includegraphics[width=.34\textwidth]{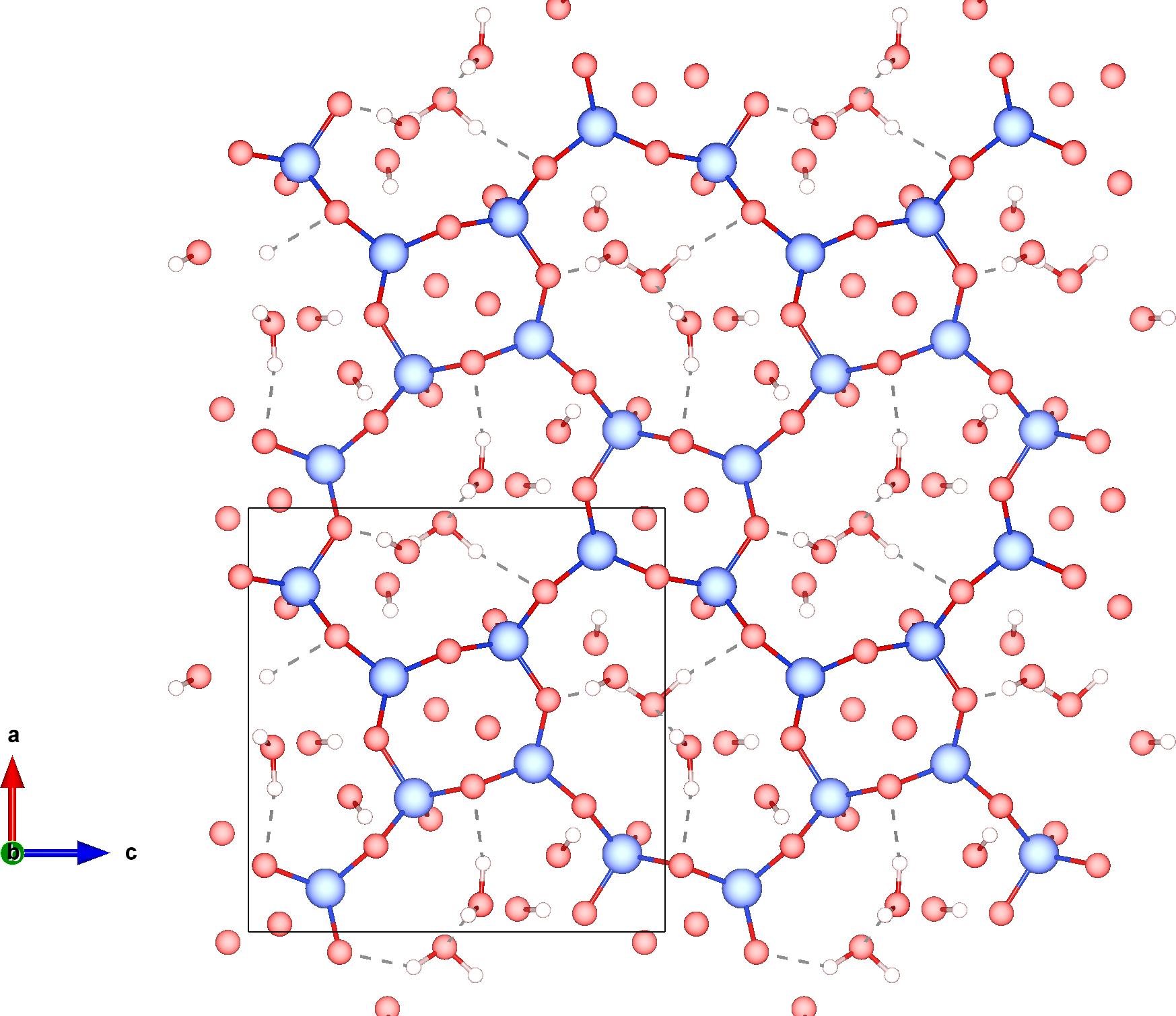}
 \includegraphics[width=.34\textwidth]{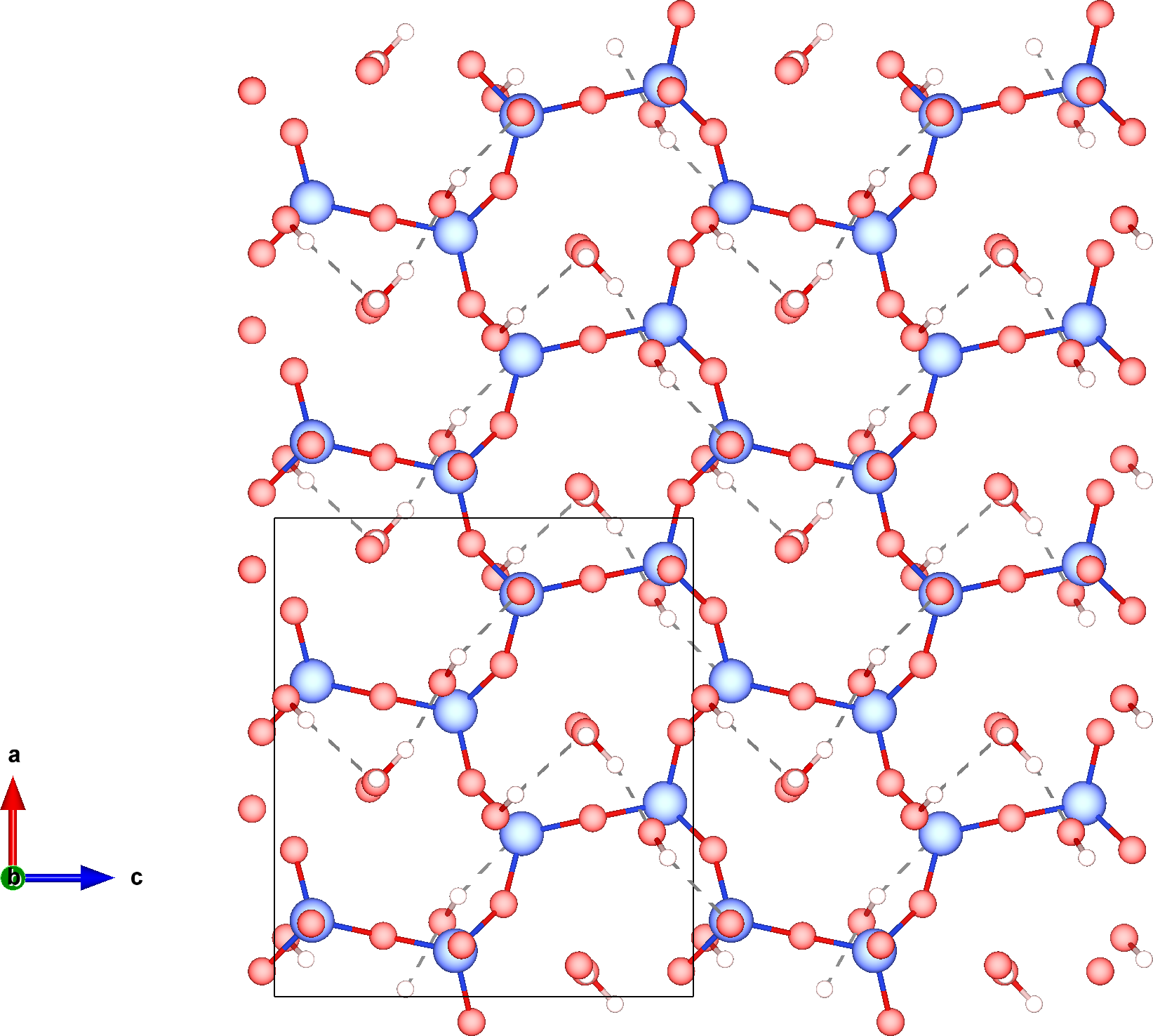}\\
 \centerline{(c) \hskip .34\textwidth (d)} ~\\
 \includegraphics[width=.34\textwidth]{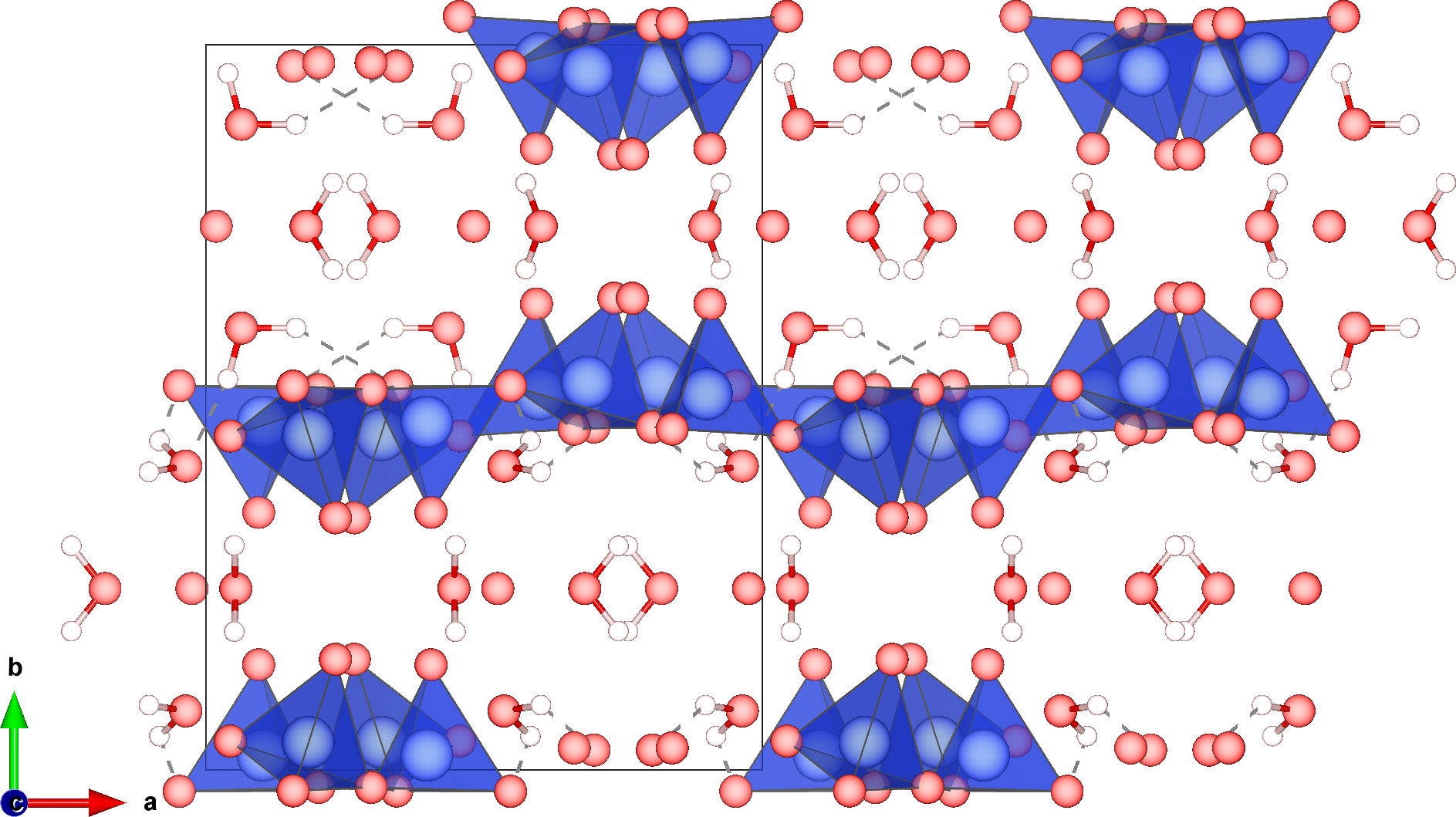}
 \includegraphics[width=.32\textwidth]{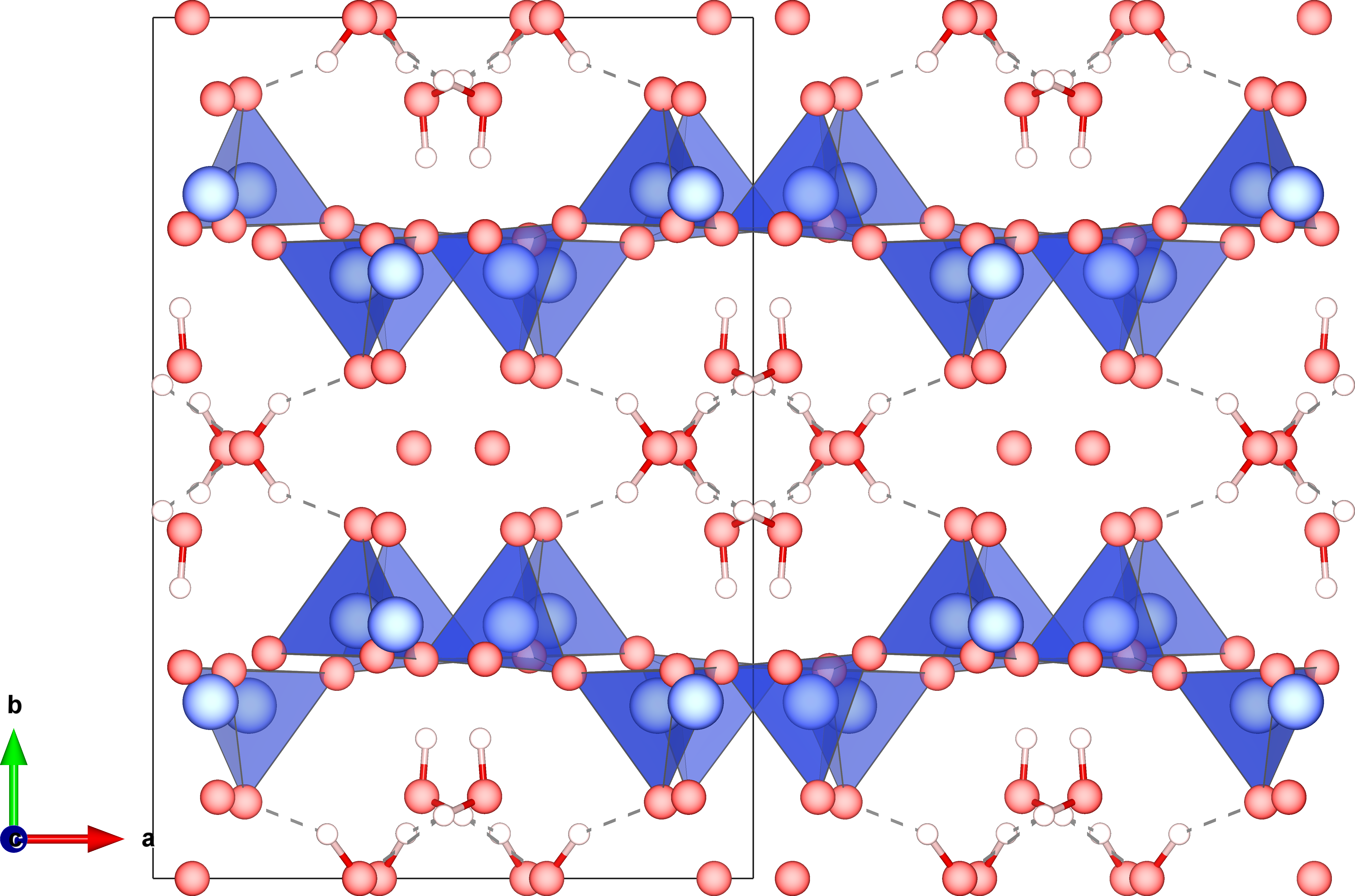}\\
 \centerline{(e) \hskip .34\textwidth (f)}
 \caption{The unit cell views of \cav\ (left panel) and \pen\ (right panel). The silicon atoms are in blue, oxygen atoms in red, calcium atoms in gray, vanadium atoms in green while hydrogen atoms are shown in white color. The polyhedra have same color themes. (a) and (b) are full unit cell structure shown at a visually convenient angle. For better understanding of underlying silicate framework we hide vanadium and calcium atoms in the following sub-figures. (c) and (d) show the `top view' (along \textbf {b} axis) without the polyhedra, highlighting the ring-like structures of silicate sheets. Similar view along the \textbf{c} axis (which includes the polyhedra for better understanding) is shown in (e) and (f) for \cav\ and \pen\ respectively. \label{fig:unitcells}}
\end{figure}

\begin{figure}
 \centering
 \includegraphics[width=.7\linewidth]{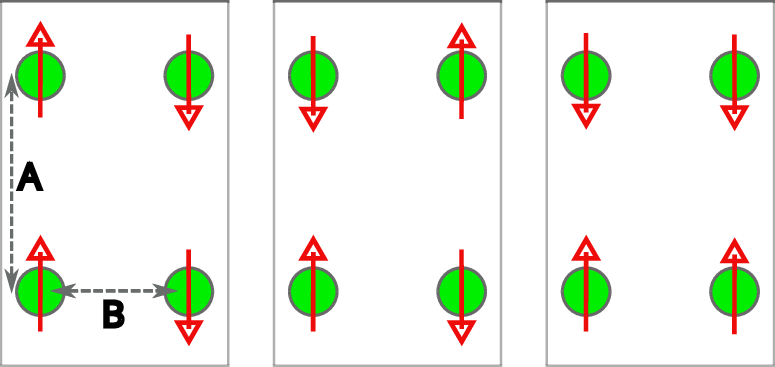}\\
 \centerline{(a)\hskip .2\textwidth (b) \hskip .2\textwidth (c)}
 \caption{A simplistic schematic of three types of antiferromagnetic configurations. The four vanadium atoms are arranged in a parallelogram of sides $A$ and $B$. For \cav\ $A=6.88${\AA} and $B=5.54$\AA, while for \pen\ $A=8.82${\AA} and $B=4.42$\AA.}
 \label{fig:magnetismschematics}
\end{figure}

\begin{figure}
 \centering
 \includegraphics[width=0.49\linewidth]{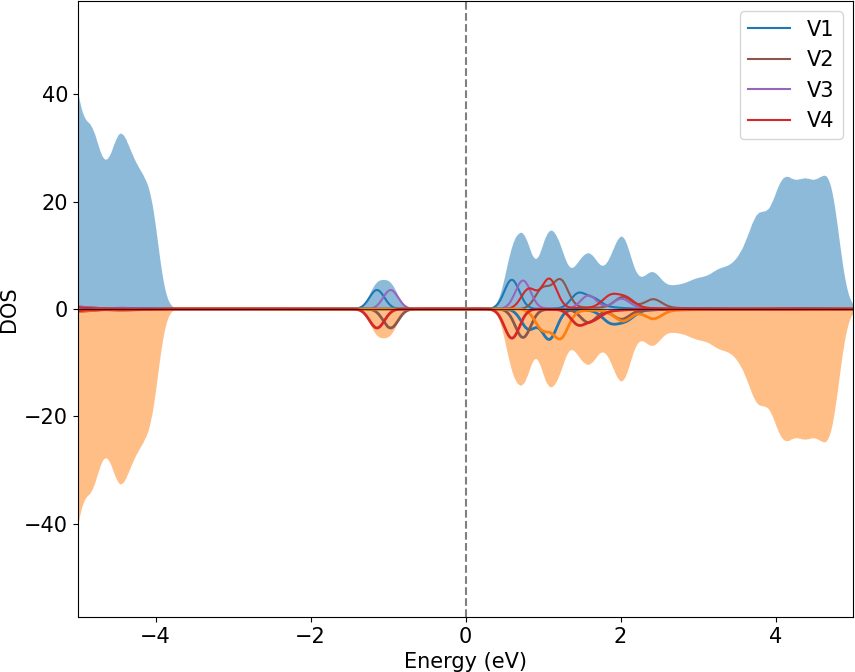}
 \includegraphics[width=0.49\linewidth]{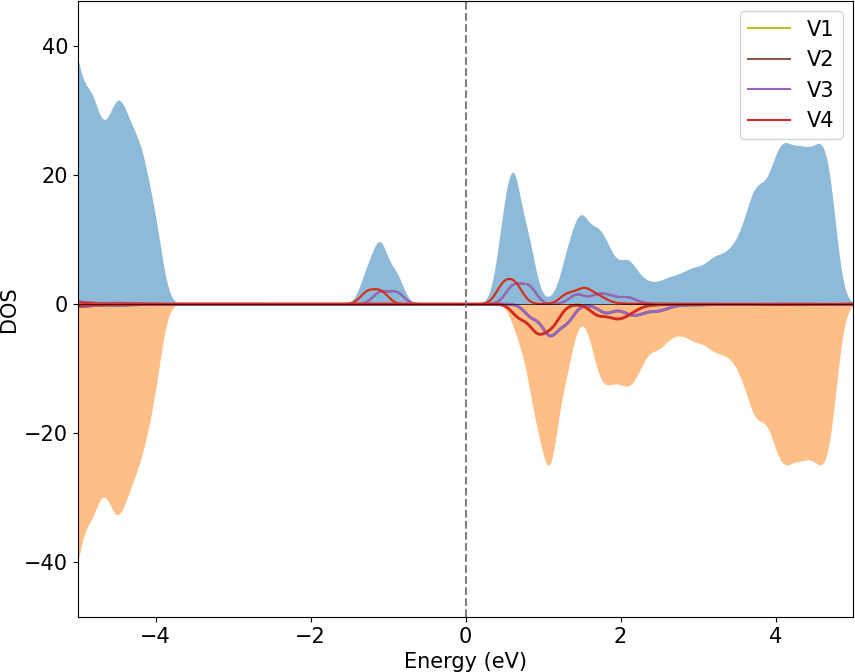} \\
 (a) \hskip .5\textwidth (b) \\~\\
 \includegraphics[width=0.49\linewidth]{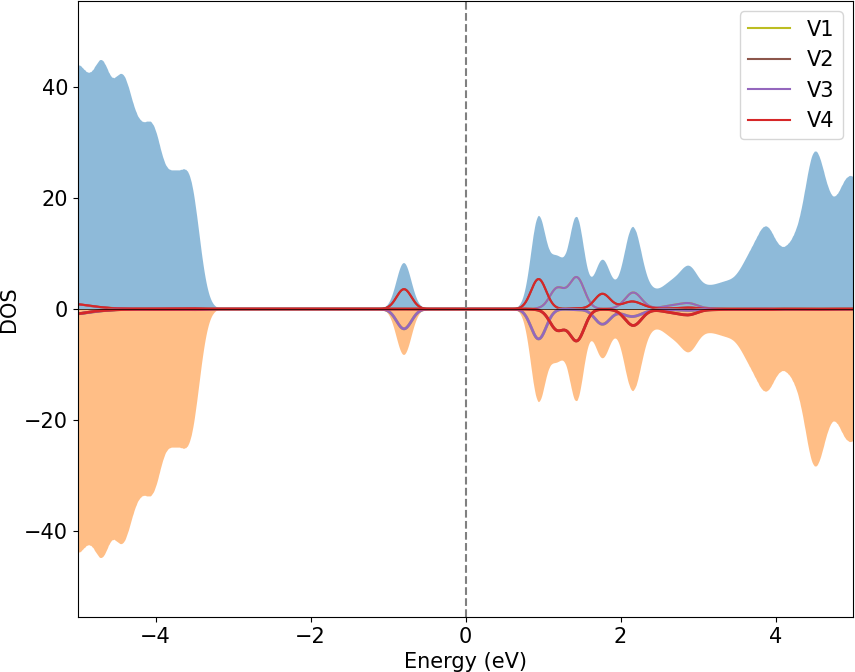}
 \includegraphics[width=0.49\linewidth]{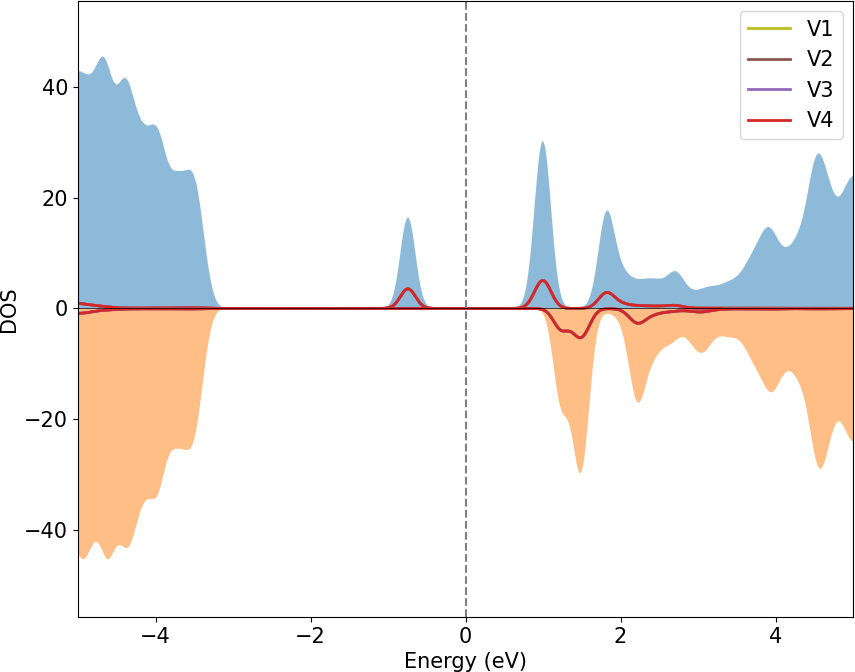}\\
 (c) \hspace {.5\textwidth} (d) \\
 \caption{Total (shaded) as well as projected (lines) density of states for \cav; (a) antiferromagnetic and (b) ferromagnetic. (c) and (d): the same for \pen. For the sake of visualization the down spin channel is shown with negative values.}
 \label{fig:cavansiteaf2tdosddos}
\end{figure}

Fig.\ \ref{fig:unitcells} shows the fully-optimized unit cells of \cav\ and \pen.
These are computed using full spin-polarized calculations as warranted by the presence of a transition element; i.e.\ Vanadium.

As there are four vanadium atoms, there can be four possible magnetic configurations: one ferromagnetic and three antiferromagnetic.
Fig.\ \ref{fig:magnetismschematics} shows the schematics of the three antiferromagnetic phases.
The ferromagnetic phase (i.e., all moment pointing in the same direction) is not shown.
These four calculations have been performed by first initiating appropriate moments on the vanadium atoms and then fully relaxing the unit cell. 

\begin{table}[tp]
\begin{center}
\begin{tabular}{c|p{.2\textwidth}|p{.2\textwidth}|p{.2\textwidth}}
\hline
 & Present study & Experiment \cite{evans1973crystal} & Experiment \cite{ishida2009polymorphic} \\
 \hline
 \cav\ & $a = 9.94 $\AA \newline $b=12.94$\AA \newline $c=9.76$\AA \newline $V=1256.19\AA^3$
 & $a=9.792 $\AA,\newline $b=13.644$\AA \newline $c=9.629$\AA \newline 
 $V=1286.45\AA^3 $ &
 $a=9.794 $\AA,\newline $b=13.670$\AA \newline $c=9.643$\AA \newline 
 $V=1291.1\AA^3 $\\
 \hline
 \pen\ & $a=10.09$\AA\newline $b=14.48$\AA\newline $c=8.84$ \newline $V=1291.19\AA^3$ & 
 $a=10.386\AA$ \newline 
 $b=14.046\AA$\newline
 $c=8.975\AA$\newline
 $V= 1309.29\AA^3$ &
 $a=10.376\AA$ \newline 
 $b=14.062\AA$\newline
 $c=8.984\AA$\newline
 $V= 1310.8\AA^3$
\end{tabular}
\end{center}
\caption{Comparison of lattice parameters obtained from the present study and available experimental measurements.}
\label{tab:lattic}
\end{table}%

For both the minerals, the optimized unit cells for all the four types of configurations are numerically identical up to three decimal places, although their energies differ as will be seen later. The unit cells are orthorhombic with the cell parameters as given in the Table \ref{tab:lattic}.
The table also shows the comparison of results of the present study with two experimental results.
That the \pen\ unit cell is 2.7\% larger in volume than \cav\ is consistent with the experimental findings.
The ground state for both the minerals is an antiferromagnetic phase depicted in Fig.\ \ref{fig:magnetismschematics}(a),  where localized moments on the vanadium atoms arranged in such a way that the nearest vanadium atoms are antiferromagnetically coupled while the next-nearest ones are ferromagnetically coupled.
The magnetic moment on each atom is approximately $0.667\mu_B$ due to partial occupancy of $d$-orbitals.
The electronic density of states are shown in Fig.\ \ref{fig:cavansiteaf2tdosddos} together with the contributions of the vanadium atoms.
The left panel of the figure shows the density of states corresponding antiferromagnetic phase of \cav\ (top) and \pen\ (bottom), and the right panel shows the same for ferromagnetic phase.
The reader may notice highly localized $d$-states at $\sim 1.3$ eV arising from vanadium atoms as a mid-gap state in an otherwise large band gap ($\sim$ 4.6 eV) for all the cases.
The differentiating factor between the two minerals is that for \pen\ the contribution from all the four vanadium atoms is degenerate in the mid-gap state while for \cav\ the pairs of nearest neighbors are degenerate and there is a marginal shift between the two pairs. 
This can be attributed to larger separation of next nearest neighbors; i.e., 6.88{\AA} for \cav\ against 8.82{\AA} for \pen\ (see Fig.\ \ref{fig:magnetismschematics}) resulting in a weaker coupling.

Although the magnetic coupling among the vanadium atoms appears stronger in \cav\ than in \pen, it is anything but strong in absolute terms.
The energy difference between the ground state antiferromagnetic and ferromagnetic phases is barely 0.6 meV for \cav\ and 0.1 meV for \pen.
The approximate Curie temperature $T_c$ can be computed using the formula
\begin{equation}
T_c = \frac{2J_{eff}}{3zk_B} = \frac{2 \Delta E} {3zk_B}.
\end{equation}
Here, the effective exchange constant ($J_{eff}$) is taken to be the difference $\Delta E$ in total energies of the magnetic phases where $z$ is the number of contributing magnetic atoms in the system ($z=4$ in our case) and $k_B$ is the Boltzmann constant.
Using this, the $T_c$ for the magnetic phase transition turns out to be, respectively, 1.32 K and 0.26 K for \cav\ and \pen.
This suggests that that the magnetic vanadium ions interact extremely weakly with each other and, for all practical purposes, it would be difficult to experimentally detect any magnetic response in either of the minerals. 

Energetically, \cav\ is seen to be lower by 0.26 eV per formula unit than \pen.
This can be attributed to their structural differences.
Both the unit cells are characterized by sheets of Si-O$_4$ polyhedra in the \textbf{a}-\textbf{c} plane, stacked along the \textbf{c} axis.
Also, the Si-O$_4$ polyhedra form rings when viewed along the longer \textbf{b} axis. 
Four water molecules per formula unit reside among the voids created by the sheets and the intermediate atoms of calcium and vanadium.
However, a few more important differences in the structures of \cav\ and \pen\ need to be highlighted.
In \cav, calcium atoms have eight coordinating oxygen atoms: four from SiO$_4$ polyhedra and four from water.
On the other hand, in \pen\ the calcium atoms have seven oxygen atoms: four from SiO$_4$ polyhedra and only three from water, with the remaining water molecule hydrogen-bonded with other water molecules and oxygen atoms of silicate sheets.
The polyhedral rings in \pen\ are 6-membered rings while those in \cav\ are larger 8-membered rings joined via smaller 4-membered rings.
This rearrangement is also associated with the restructuring of SiO$_4$ sheets. 
The reader may notice in Fig.\ \ref{fig:unitcells} (bottom panel) that for \pen\ the polyhedra are almost parallel (and anti-parallel) to each other, but for \cav\ they are tilted alternatively by 20$^\circ$ and 22.5$^\circ$.
This restructuring enables more compact packing (hence the reduction in volume) thereby lowering the ground-state energy of \cav. 

From the ground-state energies, it is therefore reasonable to qualitatively expect that \cav\ will be more abundant than \pen.
Yet one cannot ignore the effect of pressure and temperature on their formation.
This will be investigated below. 

\subsection{Temperature-dependent abundances}
\label{s:t-abundances}

Approximate temperature-dependent abundances of a two-state system may be modeled using the following Boltzmann probability model:
\begin{equation}
 p_i = \frac{e^{-E_i/k_BT}}{\sum_j e^{-E_j/k_BT}}, \label{eq:boltz}
\end{equation}
where $p_i$ is the probability of formation of state $i$ (with $i,j \in$ \{\cav, \pen\}) at temperature $T$; $E_i$ is the energy of the state $i$; and $k_B$ is the Boltzmann constant.
Because this is a two-state model, it follows that $p_{\mbox{\scriptsize \cav}} + p_{\mbox{\scriptsize \pen}} = 1$.

\begin{figure}
 \centering
 \includegraphics[width=0.99\linewidth]{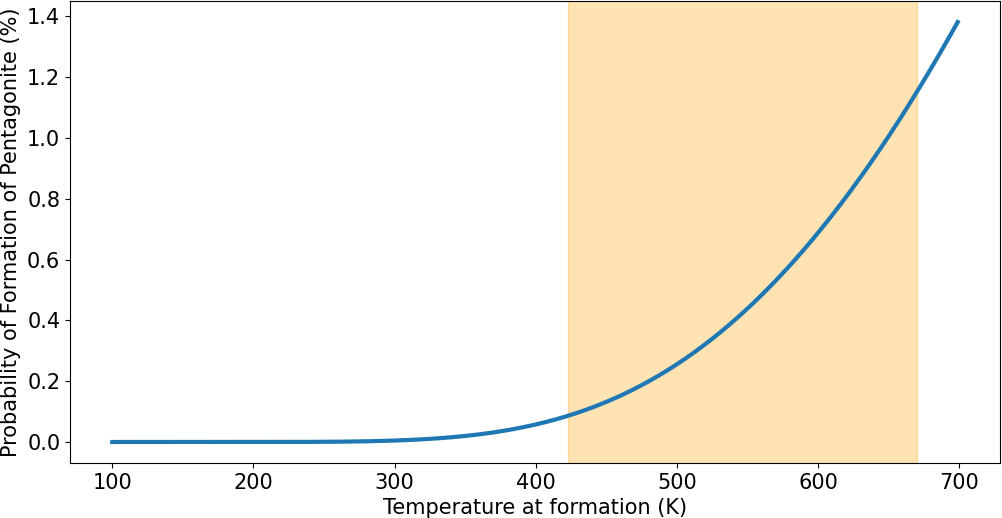}
 \caption{Probability for formation of \pen\ as a function of temperature. The shaded area indicates a geophysically plausible range of crystallization temperatures.}
 \label{fig:probofpentagonite}
\end{figure}

The probability of formation of \pen\ at zero pressure, computed as described, is shown in Fig.\ \ref{fig:probofpentagonite}.
As of today, it is not clear what the crystallization temperatures for \cav\ and \pen\ in lava flows are, but as reported \cite{danisi2012situ}, \cav\ maintains crystallinity even when heated up to 670 K.
The lava temperature at the time of formation is expected to be around or below this temperature and down to 450 K.
For visualization purposes, this plausible regime of temperatures is highlighted in the figure.
As seen in the figure, about 1\% \pen\ and 99\% \cav\ is expected at around 650 K.
In other words, for every 99 units of weight of \cav, one would find barely 1 unit weight of \pen\ formed.
These numbers might provide an insights as to why it is so difficult to find \pen\ in the field.

A few caveats are in place here.
The two-state Boltzmann probability model (Eq.\ \ref{eq:boltz}) is a fairly simplistic view of what is arguably an extremely complex system of multiple interacting atomic and chemical units in a cooling-lava non-equilibrium environment.
Specifically, this model assumes that there are no competing minerals during the formation of \cav\ and \pen.
While atomistic modeling of the cooling of a lava flow is practically impossible, it may be noted that the two-state Boltzmann probability model offers a useful first approximation.
 
\begin{figure}
 \centering
 \includegraphics[width=\columnwidth]{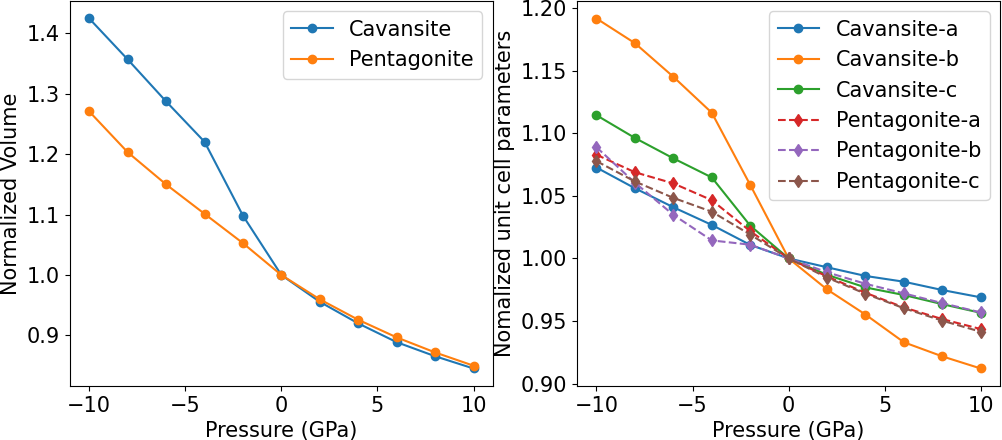}
 \caption{Evolution of (a) volume and (b) lattice parameters for \cav\ and \pen\ as a function of pressure. Unit cell parameters for \pen\ are shown by dashed lines to distinguish them from those of \cav. Note the somewhat drastic variations seen for \cav\ as compared to those for \pen.}
 \label{fig:anisopress}
\end{figure}

\begin{figure}
 \centering
 \includegraphics[width=0.99\linewidth]{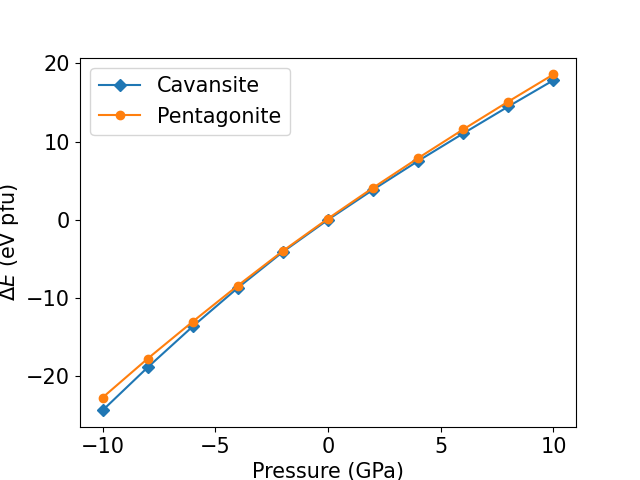}
 \caption{Variation of energies with pressure for both the systems. The zero is taken to be the ground state of \cav\ at zero pressure.}
 \label{fig:pressurevsenergy}
\end{figure}

\subsection{Pressure-dependent abundances}
\label{s:p-abundances}

To examine the effect of pressure on the formation of \cav\ and \pen, we optimize the unit cells 
(anisotropically) at a specific target 
pressure. This is done by setting the target pressure and letting system vary the cell parameters and atomic positions so as to minimize the enthalpy. The enthalpy $H$ is calculated at given pressure $P$ and volumes $V$ as: \begin{equation}
H= E  + PV,
\end{equation} where $E$ is the total energy.

Fig.\ \ref{fig:anisopress} presents the results of these optimizations including modified unit cell parameters and volumes normalized with respect to the ground state (i.e, at zero pressure).
It illustrates the effect of pressure from -$10$ GPa to $+10$ GPa.
It is worth noting that negative pressure (i.e., tensile strain or stretching of the unit cell) is of pure theoretical interest and is unlikely to be observed under geological conditions.
It is evident from the results presented that the lattice structure of \cav\ experiences a significant change primarily driven by the variations along the \textbf{b} axis.
Variations in unit cell of \pen\ are relatively monotonous.
This observation is consistent with the experimental finding by Danisi et al \cite{danisi2015elastic} (see Fig.\ 5 of the reference) who have noted similar anisotropic compressibility along the \textbf{b} axis only for \cav. 

The fact that \pen\ displays less compressibility along all axes suggests a high bulk modulus.
The bulk modulus is computed by fitting the Birch-Murnaghan equation-of-state \cite{PhysRev.71.809} to the volume-energy data. 
Fig.\ \ref{fig:pressurevsenergy} shows the energies (with respect to the ground state energy of \cav\ at zero pressure) as functions of volume.
From these data, bulk modulii are estimated to be 33.7 GPa and 41.1 GPa for \cav\ and \pen\ respectively.
These numbers are in excellent agreement with those observed experimentally (38 GPa and 49 GPa respectively) \cite{danisi2015elastic}.
Importantly, for all positive pressures, \cav\ remains lower in energy (hence favored).

\subsection{Pressure-temperature (P-T) phase diagram of abundances}
\label{s:pt-phase-diagram}

\begin{figure*}
 \centering
 \includegraphics[width=0.9\linewidth]{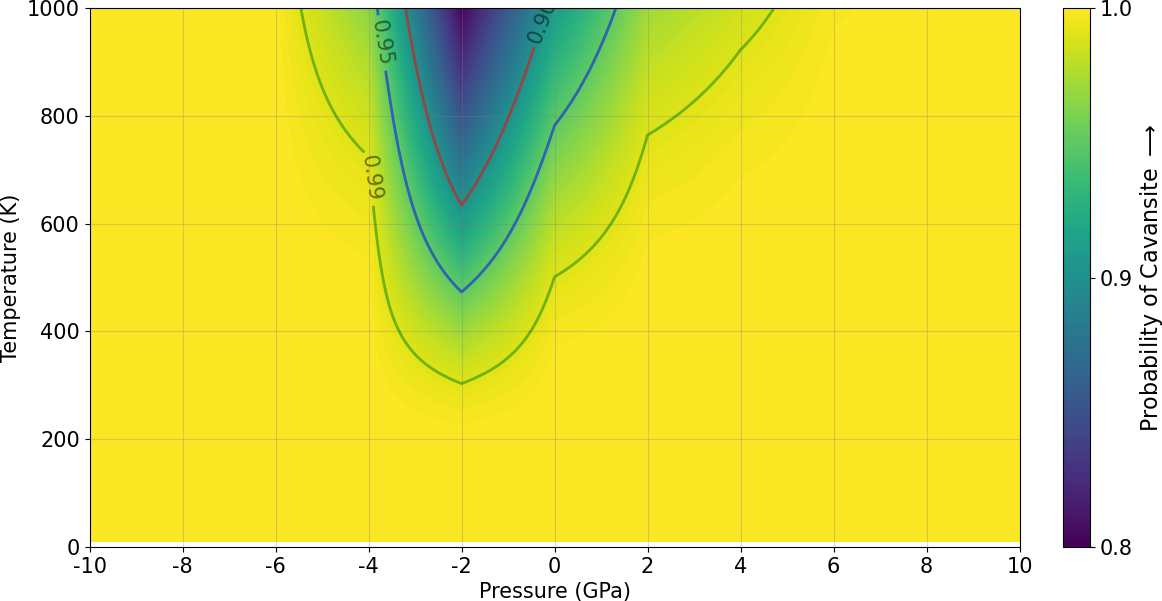}
 \caption{Probability of formations of \cav\ and \pen\ (see the color axis) as a function of pressure (x axis) and temperature (y axis). Yellow color indicates the 100\% probability of formation of \cav. Three contours corresponding to 99\% (green), 95\% (blue) and 90\% (red) formation probability of \cav\ are shown as visual aid.}
 \label{fig:prob_formation}
\end{figure*}

Sec.\ \ref{s:t-abundances} and \ref{s:p-abundances} examine the effects of pressure and temperature separately.
Here, they are combined into a P-T phase diagram.
 Using Eq \ref{eq:boltz}, we compute probabilities for a range of temperatures and pressures for both the systems. 
The resulting phase diagram is shown in Fig.\ \ref{fig:prob_formation}.
Here, colors indicate the probabilities of formation of the two species, and contours corresponding to 99\%, 95\% and 90\% probabilities of formation of \cav\ are overlaid on the plot as visual guide.
We see that \cav\ dominates the phase diagram for most part.
Because probabilities $p_{\mbox{\scriptsize \cav}}+p_{\mbox{\scriptsize \pen}}=1$ in our model (Sec.\ \ref{s:t-abundances}), these contours also indicate the 1\%, 5\% and 10\% probabilities of formation of \pen.
Clearly, pressure and temperature both appear to aid the formation of \pen\ -- however so slightly -- with the effect becoming less and less prominent at higher pressures.
(As noted before, negative pressures are unlikely under geological conditions of formation of these minerals.)
It appears that there may be fairly narrow window in the P-T plane which favours the formation of \pen. 

Ishida et al \cite{ishida2009polymorphic} note that \pen\ is a high-temperature polymorph of \cav.
This is consistent with the results of the present analysis.
A higher probability requires even higher temperatures.
In other words, presence of \pen\ in the lava flow cavities might suggest either higher lava or fluid temperatures for substantial duration.

\subsection{Dehydration of crystals}
 
\begin{table}
 \begin{tabular}{c|c|c|c|c|c}
 %Crystal 
 ~ & \textbf{a} & \textbf{b} & \textbf{c}& $\Delta V (\AA^3)$ & Reported $\Delta V (\AA^3)$ \cite{danisi2012situ} \\
 \hline
 Cavansite &9.940 & 12.944 & 9.764 &\\
 25\% \dcav&9.724&13.404&9.606& -4.2 & -6.19\\
 50\% \dcav& 9.577&12.782&9.472 & -96.8 & -82.03\\
 75\% \dcav& 9.309&12.557&9.358 &-162.3 & -132.92\\
 Dehydrated \cav\ &8.960&12.102&9.046&-275.4 & \\
 \hline
 Pentagonite &10.093&14.475&8.838& &\\
 25\% \dpen&10.034&14.006&8.850& -47.8& \\
 50\% \dpen& 10.026&13.863&8.905& -53.6&\\
 75\% \dpen& 9.932&13.256&8.726 & -142.4&\\
 Dehydrated \pen& 10.012&12.318&9.032 & -177.3&\\
 \end{tabular}
 \caption{Cell parameters of \cav\ and \pen\ at different values of dehydration. The change in volumes ($\Delta V$) with respect to the fully hydrated cell is also shown. Additionally last column shows the experimentally available values of $\Delta V$.}
 \label{tab:cell}
\end{table} 

%\begin{figure}
% \centering
% \includegraphics[width=0.9\linewidth]{fig8.png}
% \caption{On the left axis: Variation of energy at different levels of hdration with respect to 100\% hydrated state of \cav\ (blue) and \pen\ (red). We note that the two lines are almost overlapping. On the right axis: Experimentally reported temperatures of dehydration of \cav. (Note the Y-axis on the right is reversed to aid the visualization.) \cite{ishida2009polymorphic}.}
% \label{fig:deltaevshydration}
%\end{figure}

The role of water molecules in the structure is now investigate by removing one water molecule at a time and then performing full unit cell relaxations.
For each level of (de)hydration, multiple different unit cell configurations are possible.
Specifically, for $k$ water molecules per unit cell, there can be ${4 \choose k}$ unit cell configurations ($k=0,\ldots,4$).
% $\left( \begin{array}{c}4\\k \end{array}\right)$
% $\binom{4}{k}$ % amsmath is not supported by iop 
All the configurations for each $k$ are systematically generated and fully optimized.
Reported here are only the ones with the lowest energy.
See Table \ref{tab:cell} for our optimized cell parameters and a comparison with experimentally available values\cite{danisi2012situ}.
The reader may note that the unit cell volume decreases with the removal of water.
This is expected and indicates the reduction in volume of the voids inside a unit cell.
These results are consistent with experimentally available values including the drastic reduction in \cav\ volume from 25\% to 50\%. However, note that the experimental values are obtained at significantly high temperatures -- a factor that cannot be easily incorporated in the present results.
Overall, it is seen that \cav\ undergoes a more dramatic volume loss compared to \pen.

\begin{figure*}
 \begin{center}
 \includegraphics[width=.45\textwidth]{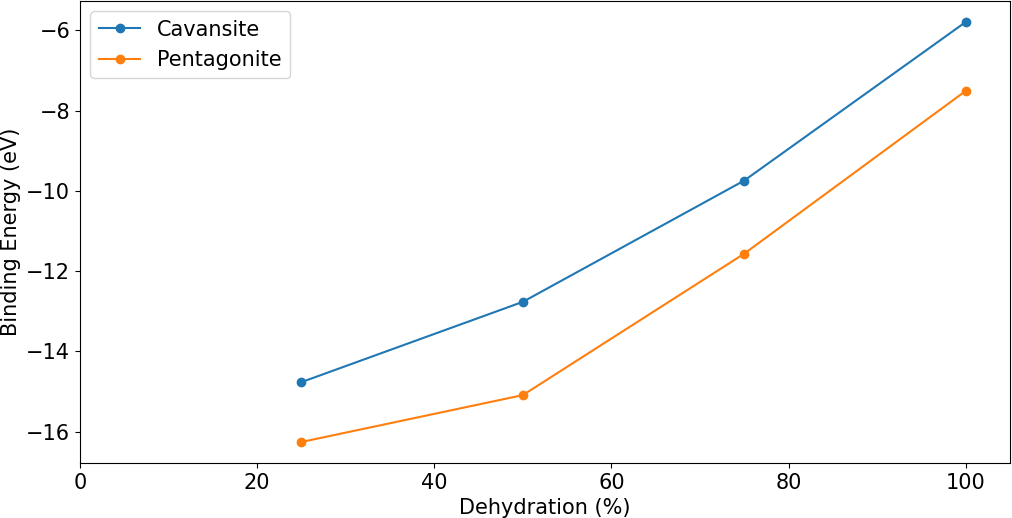}
 \includegraphics[width=.45\textwidth]{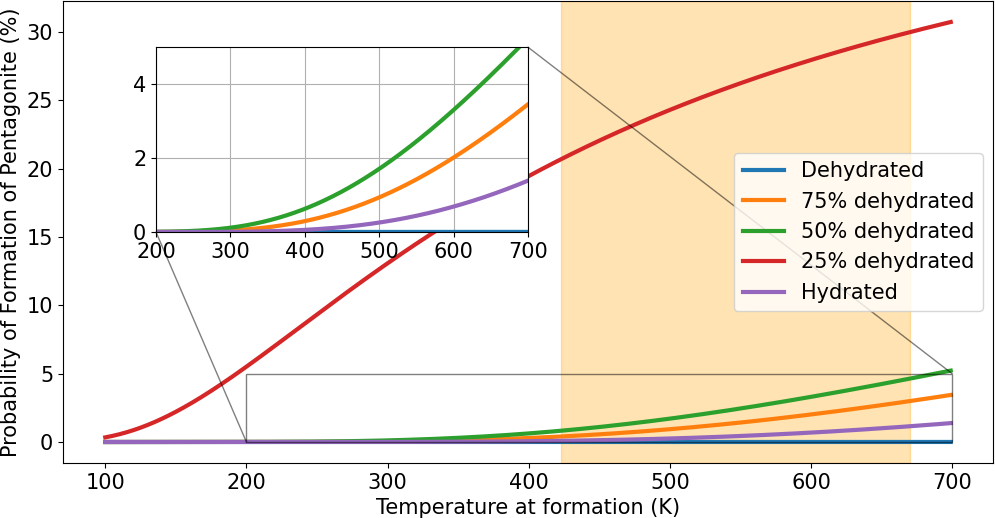}\\
 (a) \hskip .4\textwidth (b)
 \end{center}
 \caption{(a) Binding energies of water molecules as a function of dehydration. We note relative flattening of the \pen\ curve (orange) at the lowest value of dehydration signifying the relatively less resistance for the removal of water. (b) Relative abundance for \pen\ as a function of dehydration.}
 \label{fig:BE_abund}
\end{figure*}

Fig.\ \ref{fig:BE_abund}(a) shows binding energies of water at individual step of dehydration.
Although \cav\ as a whole has overall lower energy, the individual water molecules have less binding energy than those in \pen.
This is true irrespective of the level of dehydration.
We also see nearly monotonous reduction in binding for \cav\ upon increasing levels of dehydration.
It is to be noted that \pen\ shows a slight flattening of the curve at 25\% of dehydration.
This is because, as explained before, calcium in \cav\ forms octahedra which harbour all four water molecules, in contrast to \pen\ for which one of water molecule is not bound to calcium octahedra.
As a result, it is relatively easy to remove the first water molecule in \pen than other rest three.
This amounts to reduction in the overall energy difference between \cav\ and \pen\ which is further reflected in the abundance curves (Fig.\ \ref{fig:BE_abund} (b)).
As seen from the figure -- except for 25\% dehydration -- the probability of formation of \pen\ follows a similar exponential trend as in Fig.\ \ref{fig:probofpentagonite}.
However, the probability of formation increases with dehydration.
This increase is drastic at 25\% dehydration due the reason mentioned above.
Overall, one might conjecture that water-stressed environment  might lead to a higher abundance of \pen.
Yet, there is no evidence from the field data that indicates any insufficient inclusion of water molecules in these crystals.
Samples reported in the literature seem to indicate complete hydration.

At this point, the role of van der Waals (vdW) interaction is worth a note .
Without vdW corrections, it is known that DFT calculations overestimate typical cell lengths by about 10\% compared to uncorrected calculations.
This is, of course, expected because non-vdW-corrected DFT functionals do not accurately capture the long-range attractive vdW interaction.
In the present context, it is seen that in absence of the vdW correction, the ground state energy of dehydrated \pen\ is lower than that of dehydrated \cav.
On the other hand, vdW-corrected calculations predict that \cav\ is the energetically more favourable ground state structure.
A conclusive resolution of this issue may be possible only through experiments.

\section{Implications of our results to mineralisation}

The computational analysis and results presented in this paper may have interesting implications to the current understanding of the mineralisation processes leading to the formation of \cav\ and \pen. This is elaborated upon below. 

Wagholi, Pune, India, is the seat of currently the largest deposits of both \cav\ and \pen.
Here, a thick ($\sim$10 m), massive basaltic lava flow is exposed and is quarried for road metal and aggregates.
At its base, its contact with the lower lava flow is marked by sporadic patches of flow-top breccia.
This highly oxidized, glassy breccia has high porosity and permeability with copious irregular and haphazard vugs and cavities containing heulandite, stilbite, a rare green apophylite, and two generations of calcite.
Radiated aggregates or sheaf-like crystals of \cav\ grow on top of heulandite, stilbite and calcite.
In 1998, the rarer \pen\ was discovered to co-occur with its diamorph \cav\ in some of the cavities \cite{ottens2000cavansit}.
\cav\ and \pen\ occurrences at Wagholi are characterized by high vanadium concentrations of 600-750 ppm in the host basalts \cite{ghodke1984trace}.
Thus, the presence of vanadium-bearing minerals suggests in situ scavenging of vanadium from glassy basalt by hydrothermal fluids and their subsequent deposition in micro niche environments \cite{ottens2003minerals}.

The findings of the present study suggest the following possibilities about the mineralisation processes that lead to the formation of \cav\ and \pen. 
\begin{enumerate}
 \item
 It is speculated that \cav\ may have been deposited first by hydrothermal fluids under saturated conditions (which is a common process for zeolite formation in cavities over relevant geological time scales).
 \pen\ got deposited only after the temperature in the cavities rose and evaporated the fluid in the cavities at slightly faster rates (explaining relatively small crystal sizes).
 The top layer of lava mentioned earlier could be one possible reason for temperature rise in the lower layer because of blanketing of the lower layer. 

 \item
 The observed abundances of \cav\ and \pen\ are explained through two mechanisms which may act separately or in tandem under geological conditions:
 \begin{enumerate}
  \item
  Through the fundamental physics of these two minerals leading to the probabilities presented earlier ($\sim$1\% for \pen\ and $\sim$99\% for \cav\ in the relevant lava temperature range of 450-650 K.
  It is speculated that the thick lava flows from Wagholi might have cooled under a similar thermal regime (perhaps even lower) when zeolite mineralisation was initiated in the breccia cavities. 

  One must also remember that breccia are known to experience considerable overburden pressure in addition to the fluid pressure in the cavities.
  In the P-T phase diagram (Fig.\ \ref{fig:prob_formation}), there is a distinct area around 4 GPa at high enough temperatures where we see a significant probability of formation of \pen.
  However, these conditions are rarely achieved in real-life lava flows.
  Consequently, the probability of formation of \pen\ remains significantly low despite higher temperatures.

  \item
  Via the normal geological spatiotemporal variability of prevailing temperature and pressure conditions, suggesting that only a limited fraction of vugs and cavities may have been exposed to appropriate temperature, pressure and chemical conditions favouring the formation of \pen.
 \end{enumerate}
\end{enumerate}

 \section{Conclusion}

 In this work, extensive DFT calculations have been performed to examine the electronic structures, structural properties, and relative abundances of two microporous vanadosilicate dimorphs, namely \cav\ and \pen.
 This analysis finds that the structural arrangements of SiO$_4$ tetrahedra as well as the differently-coordinated calcium atoms are responsible for the lower ground state energy of \cav\ at zero pressure and zero temperature.
 The analysis also computes electronic structures under pressure to mimic field conditions in the lava flows where these minerals are formed.
 A two-state Boltzmann probability model is used to model the effect of temperature on the relative abundances of the two minerals.
 This leads to the construction of a comprehensive pressure-temperature-abundance phase diagram for the two minerals.
 This phase diagram explains why \cav\ is significantly more abundant than \pen.
 This analysis further suggests that the formation of \pen\ is assisted by both pressure and temperature although this effect is most pronounced over a limited range of (negative) pressures.
 This study provides insights into mineralization processes that lead to or contribute to the deposition of \cav\ and \pen\ in the observed proportions.
 The DFT calculations presented here also predict a highly localized weak magnetic state due to vanadium atoms that is associated with weak magnetic ordering at very low temperatures (Curie temperatures around 1 K).
 Hydration in both the minerals is investigated, and indicates that water molecules are somewhat loosely bound to the crystal structure.
 This work sheds light on both for the geological and material science aspects of vanadosilicate minerals and paves way for further studies of their formation and properties.

%\begin{acknowledgement}
\ack
 BSP thanks Andrey Tokarev for valuable comments. 
%\end{acknowledgement}

% \paragraph{Funding Statement}
% This research was supported by grants from the <funder-name> <doi> (<award ID>); <funder-name> <doi> (<award ID>).

\paragraph{Competing Interests}

None

%\endnote in some journals will behave like \footnote; and \printendnotes will not output anything. 
%\printendnotes

%\printbibliography
\section*{References}
\bibliography{biblio}

\appendix
\section*{Appendix}
\setcounter{section}{1}

%\begin{table*}[]
 \begin{longtable}{@{}p{.2\textwidth}|p{.18\textwidth}|p{.15\textwidth}|p{.2\textwidth}|p{.22\textwidth}@{}}
 \toprule
 Locality &
 Lava type &
 Host locale &
 Mineral association &
 Reference \\ \midrule
 \multicolumn{5}{l}{\textbf{Columbia River Flood Basalt Province, USA}} 
 \\ \midrule
 Qwyhee Dam, Oregon &
 Rhyolite, tuff breccia intruded by dyke &
 Red tuff breccia &
 \cav, heulandite, stilbite, calcite, apophyllite, analcime, 
 rare pentagonite, native copper &
 Staples et al.\cite{staples1973cavansite} \\ \midrule
 Charles W. Chapman quarry, Goble, Oregon &
 Rubbly pahoehoe intruded by dykes &
 Cavity filling, calcite veins &
 \cav, calcite, heulandite, thomsonite, native copper &
  Staples et al.\cite{staples1973cavansite} \\ \midrule
  \multicolumn{5}{l}{\textbf{Deccan traps, India}} 
  \\ \midrule
 Wagholi, Pune &
 Rubbly pahoehoe &
 Flow top breccia (FTB) &
 \cav, Pentagonite, stilbite, mordenite, heulandite, calcite, 
 apophyllite, native copper &
 Evans \cite{evans1973crystal}, Birch \cite{birch1977cavansite}, 
  Kothvala \cite{kothavala1991wagholi}; Cook 
 \cite{cook1996connoisseur}, 
 Blass et 
 al. \cite{ottens2000cavansit}, Powar and 
 Byrappa\cite{powar2001x}; 
 Ottens \cite{ottens2003minerals}, 
  Makki \cite{makki2005collecting} \\ \midrule
 Sutarwadi, Pune &
 Sheet pahoehoe &
 Gas blister &
 \cav, stilbite &
  Mookherjee and Phadke \cite{mookherjee1998thermo} \\ \midrule
 Yedgaon Dam, Narayangaon &
 Sheet pahoehoe &
 Cavity filling &
 \cav\ &
  Mookherjee and Phadke \cite{mookherjee1998thermo} \\ 
 \midrule
 Surli Ghat, Karad &
 Rubbly pahoehoe &
 FTB &
 \cav\ &
  Mookherjee and Phadke \cite{mookherjee1998thermo} \\ \midrule
 Lonavala Quarry &
 Sheet pahoehoe &
 Cavity filling &
 \cav, calcite, heulandite-Ca, modernite, pentagonite, quartz &
  Praszkier and Siuda, \cite{praszkier2007lonavala} \\ 
 \midrule
 Indori, Talegaon &
 Sheet pahoehoe &
 Cavity filling &
 \cav, stilbite, apophyllite, mesolite, modernite, heulandite, 
 calcite &
  Praszkier, \cite{praszkiercavansit} \\ \midrule
  \multicolumn{5}{l}{\textbf{ Rio Grande do Sul, Brazil}} \\ \midrule
 Morro Reuter, Brazil &
 Vesicular basalt &
 Vein filing &
 \cav, quartz &
 Frank et al., 2004 \cite{frank2004cavansit} \\ \midrule
  \multicolumn{5}{l}{\textbf{New Zealand}} \\ \midrule
 Aranga Quarry, Dargaville &
 likely Rubbly pahoehoe &
 Red breccia &
 \cav, chabazite-Ca, calcite, okenite, native copper, 
 chrysocolla &
  Frank et al \cite{frank2004cavansit} \\ \bottomrule

 \caption{Locations and details of the conditions where \cav\ and \pen\ 
 are found across the world. There are only a handful of basaltic planes 
 where they are found with the best known specimens emerging from Pune, 
 India.}
 \label{tab:locations}
 \end{longtable}
%\end{table*}

\end{document}